\documentclass[%
superscriptaddress,
reprint,
showpacs,
amsmath,amssymb,
aps,
pre,
]{revtex4-1}

\usepackage{graphicx}
\usepackage{dcolumn}
\usepackage{bm}
\usepackage{color,verbatim}

\begin {document}


\title{Anomalous diffusion in a quenched-trap model on fractal lattices}

\author{Tomoshige Miyaguchi}
\email{tmiyaguchi@naruto-u.ac.jp}
\affiliation{%
  Department of Mathematics Education, Naruto University of Education, Tokushima 772-8502, Japan
}
\author{Takuma Akimoto}
\affiliation{%
  Department of Mechanical Engineering, Keio University, Yokohama 223-8522, Japan
}%


\date{\today}


\begin{abstract}
  Models with mixed origins of anomalous subdiffusion have been considered
  important for understanding transport in biological systems. Here, one such
  mixed model, the quenched trap model (QTM) on fractal lattices, is
  investigated. It is shown that both ensemble- and time-averaged mean square
  displacements (MSDs) show subdiffusion with different scaling exponents, i.e.,
  this system shows weak ergodicity breaking. Moreover, time-averaged MSD
  exhibits aging and converges to a random variable following the modified
  Mittag--Leffler distribution. It is also shown that the QTM on a fractal
  lattice can not be reduced to the continuous-time random walks, if the
  spectral dimension of the fractal lattice is less than 2.
\end{abstract}

\maketitle


\label{sec.intro}

Anomalous diffusion has received much attention in recent years \cite{barkai12,
  *hofling13, metzler14}, because it has been reported in many single-particle
tracking experiments in biological systems \cite{golding06, *jeon11, *burov11}
and molecular dynamics simulations \cite{akimoto11, *uneyama12, *jeon12}.  In
particular, much effort has been devoted to theoretical studies to elucidate
what kind of anomalous diffusion is consistent with these experiments
\cite{golding06, akimoto11, he08, *lubelski08, *neusius09, *massignan14,
  *thiel13, *thiel14, *schulz14}. Among these studies, models with mixed origins
of anomalous diffusion are found to agree well with some experimental data. For
example, continuous time random walks (CTRWs) on fractal lattices well reproduce
the diffusion of potassium channels on plasma membrane \cite{weigel11, meroz10},
and a mixed model of fractional Brownian motion (FBM) and CTRW well explains the
diffusion of insulin granules in cells \cite{tabei13} as well as molecular
dynamics simulations of water molecules on the membrane surface
\cite{yamamoto14}. However, these theoretical models are almost
phenomenological, and their underlying mechanisms still remain to be elucidated.


As biological origins of these mechanisms, fractal structures are considered to
be generated by molecular crowding \cite{bancaud09, weigel11}; the FBM is
believed to be due to viscoelasticity of the cytoplasm \cite{tabei13}. On the
other hand, energetic disorder due to transient traps to binding sites is
considered to generate the CTRW dynamics \cite{weigel11, tabei13}. In fact, such
energetic disorder is a physical origin of CTRWs for the case of non-fractal
lattices with the spacial dimension larger than 2 \cite{machta85,
  bouchaud90}. However, little is known about such reduction for the diffusion
on fractal geometry. Here, we study random walks in random energy
landscape---quenched trap model (QTM) \cite{machta85, bouchaud90, burov07,
  *burov12, miyaguchi11b}---on fractal lattices, and show that the reduction to
CTRWs is impossible for the system with the spectral dimension lower than 2.
The results in this Rapid Communication are a generalization of
Ref.~\cite{miyaguchi11b}, which studied the QTM on hypercubic lattices.


We consider random walks on a fractal lattice with fractal dimension $d_f$ and
spectral dimension $d_s$. The fractal dimension $d_f$ characterizes a static
property (the configuration of the lattice points), while the spectral dimension
$d_s$, originally defined through the spectral density of state, characterizes a
dynamic property.
Even the discrete time random walks (DTRWs) on fractal lattices show the
anomalous diffusion \cite{havlin02}. We denote the position of the DTRW at time
$n$ by $\bm{\tilde{r}}(n) \in \mathbb{R}^d$. Then, the ensemble-averaged mean
square displacement (EMSD) of DTRWs on fractals is given by
\begin{equation}
  \label{e.eamsd.dtrw}
  \left\langle \delta \bm{\tilde{r}}^2 (n) \right\rangle \sim n^{\beta}, 
\end{equation}
where $\delta \bm{\tilde{r}} (n) \equiv \bm{\tilde{r}}(n) - \bm{\tilde{r}}(0)$,
and $\beta \in (0,1]$ (anomalous subdiffusion). The bracket $\left\langle \cdots
\right\rangle$ stands for the ensemble average over both thermal history and
random environment \cite{havlin02}.
Furthermore, the spectral dimension $d_s$ is related to the number of
visited sites until time $n$, $S_n$, as
\begin{equation}
  \label{e.S_n}
  \left\langle S_n \right\rangle \sim n^{d_s/2},
\end{equation}
and thus $d_s \leq 2$.  For the hypercubic lattice ($d_f = 1,2,3,\dots$), $d_s =
1 $ if $d_f = 1$ and $d_s = 2$ if $d_f = 2, 3,\dots$. (More precisely, a
logarithmic correction appears in Eq.~(\ref{e.S_n}) for $d_f = 2$. See
\cite{bouchaud90, miyaguchi11b} for details).


In the DTRW on fractal lattices stated above, all the lattice points are
energetically identical, while the QTM is the diffusion model on random
potential landscapes. Because the QTM is the continuous time model, we denote
the particle position on the fractal lattice as $\mbox{\boldmath $r$} (t) \in
\mathbb{R}^d$. In the QTM, a particle that arrives at a site $k$ is trapped at
that site for a time $\tau_k$ before jumping again. The trap time $\tau_k$ is
assumed to follow a power law
\begin{eqnarray}
  \label{e.trapping_dist}
  p (\tau) \simeq \frac {c_0}{\tau^{1+\mu}},
  ~~{\rm as}~~\tau \to \infty,~~(0 <\mu< 1)
\end{eqnarray}
where $\mu$ is the stable index. Also, $c_0$ is defined by $c_0 = c /
|\Gamma (-\mu)|$, where $c$ is the scale factor and $\Gamma (-\mu)$ is the
Gamma function.  We assume that the trap time $\tau_k$ of the site $k$ is
the same for each visit to this site, i.e., the random trap time $\tau_{k}$
is a quenched disorder. The origin of the power law trap time
distribution [Eq.~(\ref{e.trapping_dist})] is random potential landscapes
with the potential depths following an exponential distribution
\cite{bouchaud90}.
%
%
This Rapid Communication also presents numerical results for QTM on
two-dimensional Sierpinski gasket, for which exact values of $\beta$ and $d_s$
are known: $d_s = 2 \ln 3 / \ln 5$ and $\beta = 2 \ln 2 / \ln 5$
\cite{havlin02}.


On the basis of the analysis reported in \cite{miyaguchi11b}, we
approximately derive the probability density function (PDF) of the number
of jumps until time $t$, $N_t$, which is an important quantity because
$N_t$ connects the DTRW and CTRW. The following equation plays a central
role:
\begin{equation}
  \label{e.cumulative.1}
  \mathrm{Prob} \left( N_t < n\right)
  =
  \mathrm{Prob} \left( T_n > t \right),
\end{equation}
where $T_n$ is the time when the $n$-th jump occurs and is called the
$n$-th renewal time.


We start with the derivation of the PDF of $T_n$, then derive the PDF of
$N_t$ through Eq.~(\ref{e.cumulative.1}).  Let $l'_{k} \, (k=1, 2, \dots)$
denote the site index visited just after the $(k-1)$-th jump. Then, the
$n$-th renewal time $T_n$ is expressed as
\begin{equation}
  T_n = \sum_{k=1}^{n} \tau_{l'_k}.
\end{equation}
Note that the same integers can appear in the sequence of site indexes
$\{l'_k\}_{k=1,2,\dots,n}$, since the particle can visit the same site
repeatedly. Accordingly, the trap times
$\{\raise0.3ex\hbox{$\tau_{l'_k}$}\}_{k=1,2,\dots,n}$ are not mutually
independent.

To handle this interdependence between trap times, we rewrite $T_n$ as
follows \cite{bouchaud90}:
\begin{equation}
  \label{e.T_n.exact}
  T_n  = \sum_{k=1}^{S_n} N_{n,k} \tau_{l_k},
\end{equation}
where $S_n$ is the number of visited sites. The indexes of these visited sites
are denoted by $\{l_k\}_{k= 1, \dots, S_n}$, and $N_{n,k}$ is the number of
visits to the site $l_k$. Note that the same integers do not appear in the
sequence of site indexes $\{l_k\}_{k=1,2,\dots, S_n}$, and thus the the trap
times $\{\tau_{l_{k}}\}_{k=1,2,\dots,S_n}$ are mutually independent.

Here, let us approximate $N_{n,k}$ as $N_{n,k} \approx {n}/{S_n}$ (i.e., we
neglect fluctuations of the number of visits to each site). Moreover, in order
to use the generalized central limit theorem, we rewrite Eq.~(\ref{e.T_n.exact})
as
\begin{equation}
  \label{e.T_n.approx}
  T_n \approx
  \frac {n}{S_n} \sum_{k=1}^{S_n} \tau_{l_k}
  =
  \frac {n}{S_n^{1-1/\mu}}
  \frac {1}{S_n^{1/\mu}} \sum_{k=1}^{S_n} \tau_{l_k}
  \equiv a_n Y_n,
\end{equation}
where $a_n$ and $Y_n$ are defined as $a_n \equiv n/S_n^{1-1/\mu}$ and $Y_n
\equiv 1/S_n^{1/\mu} \sum_{k=1}^{S_n} \tau_{l_k}.$ By neglecting the
fluctuations of $S_n$ as $S_n \approx \left\langle S_n \right\rangle$ and using
Eq.~(\ref{e.S_n}), we further approximate $a_n$ as
\begin{equation}
  a_n \approx K n^{1+d_s (1-\mu)/(2\mu)},
\end{equation}
where $K$ is a constant. Since $\{\tau_{l_k}\}_{l=1,\dots,S_n}$ are mutually
independent, we can use the generalized central limit theorem \cite{feller71,
  bouchaud90}, and find that $Y_n$ converges to a random variable $Y$ as $n \to
\infty$ which follows the one-sided stable distribution $l_{\mu} (y)$. Thus,
$T_n$ also follows the $l_{\mu} (y)$ after a suitable rescaling. A series
expansion of $l_{\mu} (y)$ is given by \cite{feller71}
\begin{equation}
  \label{e.levy}
  l_{\mu} (y)
  =
  \frac {-1}{\pi y} \sum_{k=1}^{\infty}
  \frac {\Gamma (k \mu + 1)}{k!} (- c y^{-\mu})^{k} \sin (k \pi \mu).
\end{equation}


Next, we derive the PDF of $N_t$. First, let us define a rescaled variable
$X_t$ as
\begin{eqnarray}
  \label{e.rwsd.x_t}
  N_t = b_t X_t, \quad \text{with} \quad
  b_t \simeq \left(\frac {t}{K}\right)^{\mu / \alpha}, 
\end{eqnarray}
where $\alpha \in [(1+\mu)/2, 1]$ is defined by
\begin{equation}
  \label{e.alpha}
  \alpha = \mu + d_s (1-\mu)/2.
\end{equation}
This parameter $\alpha$ is important because it characterizes the deviation of
the QTM from the CTRW. The PDF of $X_t$ is the same as that of $N_t$ except for
the difference in the scale factor, thus we derive the PDF of $X_t$ instead of
$N_t$. By using these rescaled variables $X_t$ and $Y_n$,
Eq.~(\ref{e.cumulative.1}) can be rewritten as
\begin{equation}
  \label{e.rwsd.cumulative.2}
  \mathrm{Prob} \left( X_t < x \right)
  =
  \mathrm{Prob} \left( Y_n > x^{-\alpha/\mu}\right),
\end{equation}
where $x$ is defined by
\begin{eqnarray}
  \label{e.rwsd.x}
  x \equiv \frac {n}{b_t} = \left(\frac {t}{a_n} \right)^{-\mu/\alpha}.
\end{eqnarray}

Because $Y_n$ converges to the random variable following $l_{\mu} (y)$, the
right hand side of Eq.~(\ref{e.rwsd.cumulative.2}) tends to an integral of
$l_{\mu} (y)$ in the scaling limit $n \to \infty$ (with $x$ being fixed):
\begin{eqnarray}
  \label{e.rwsd.cumulative_levy}
  \mathrm{Prob} \left( Y_n > x^{- \alpha / \mu} \right)
  \simeq
  \int_{x^{-\alpha / \mu}}^{\infty} l_{\mu} (y) dy.
\end{eqnarray}
Thus we obtain the PDF of $X_t$ by taking derivatives of
Eqs.~(\ref{e.rwsd.cumulative.2}) and (\ref{e.rwsd.cumulative_levy}) with respect
to $x$ :
\begin{align}
  g_{\mu,\alpha} (x)
  \label{e.mod_ml}
  &=
  -\frac {\alpha}{\pi \mu x } \sum_{k=1}^{\infty}
  \frac {\Gamma (k \mu + 1)}{k!} \left (-c x^{\alpha}\right)^{k} \sin (k \pi \mu).
\end{align}
There are two remarks. The first remark is that, since $X_t$ converges to a
time-independent random variable that follows the PDF (\ref{e.mod_ml}), we have
$\left\langle N_t \right\rangle \sim b_t$. The second remark is that the above
PDF is a one-parameter extension of the Mittag--Leffler distribution (MLD)
\cite{feller71, aaronson97, *akimoto10} for which $\alpha=1$; thus, we call it a
modified MLD \cite{miyaguchi11b}. A qualitative difference from the MLD is that
$g_{\mu,\alpha} (x)$ diverges at $x=0$ as $g_{\mu,\alpha} (x) \sim
1/x^{1-\alpha}$. See \cite{miyaguchi11b} for more details.
Finally, we obtain the PDF for $N_t$ as
\begin{equation}
  \label{e.f-g}
  f_{\mu, \alpha} (n, t) \approx
  g_{\mu, \alpha} \left(x\right) \frac {dx}{dn} = 
  g_{\mu, \alpha} \left(\frac {n}{b_t}\right) \frac {1}{b_t} , 
\end{equation}
where we used Eq.~(\ref{e.rwsd.x}).

\begin{figure}[t]
  \centerline{\includegraphics[width=8.60cm]{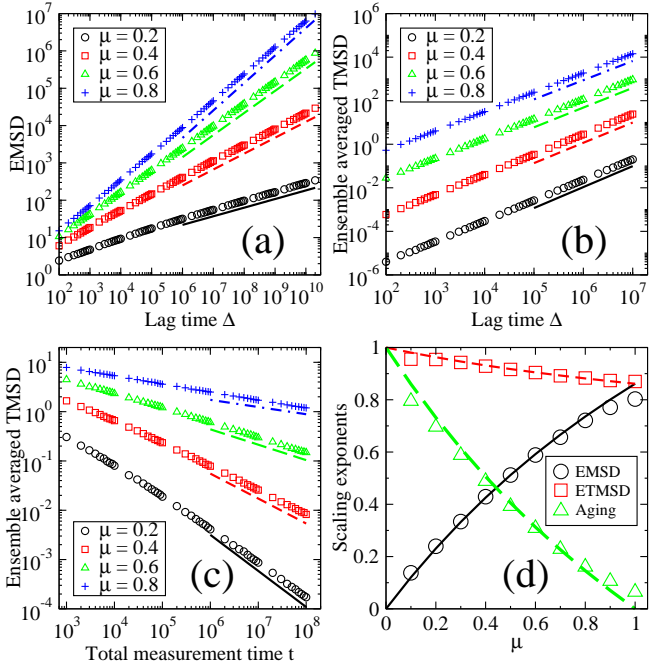}}
  \caption{\label{f.eamsd}(Color online) Symbols are numerical results for the
    QTM on the Sierpinski gasket, while curves are theoretical predictions
    [Eq.~(\ref{e.eamsd}) for the EMSD and Eq.~(\ref{e.ea-tamsd.f}) for the
    ETMSD]. The trap time PDF is set as $p(\tau) = 1 / (1 + \tau/\mu)^{\mu+1}$
    with $\tau\in (0, \infty)$. (a) and (b) The EMSD $\left\langle \delta
    \bm{r}^2 (\Delta) \right\rangle$ and ETMSD $\langle \overline{\delta
      \bm{r}^2} (\Delta, t) \rangle$ vs the lag time $\Delta$. The total
    measurement time $t$ is set as $t = 2 \times 10^{10}$. (c) The ETMSD
    $\langle \overline{\delta \bm{r}^2} (\Delta, t) \rangle$ vs the total
    measurement time $t$ [see Eq.~(\ref{e.ea-tamsd.f})]. The lag time $\Delta$
    is fixed as $\Delta = 10^2$. This figure shows aging behavior of the
    ETMSD. (d) Scaling exponents of anomalous diffusion for the EMSD (circles
    and solid line) and ETMSD (squares and dashed line) vs $\mu$. The total
    measurement time $t$ is set as $t = 2 \times 10^{13}$ for $\mu = \{0.1, 0.2,
    0.3\}$, $t = 2\times 10^{11}$ for $\mu = \left\{0.4, 0.5, 0.6\right\}$, $t =
    2 \times 10^9$ for $\mu = \left\{0.7, 0.8\right\}$, and $t = 2 \times 10^8$
    for $\mu = \left\{0.9, 1.0\right\}$. The scaling exponent of aging for the
    ETMSD is also shown (triangles and long dashed line). These scaling
    exponents are obtained by least-square fittings (under log-log form) in the
    range $\Delta \in [10^{6}, t]$ for the EMSD (circles), and $\Delta \in
    [10^{5}, 10^{7}]$ (squares) and $t \in [10^{6}, 10^{8}]$ (triangles) for
    the ETMSD.}
\end{figure}


Next, we derive asymptotic formulas for the EMSD $\langle \delta \bm{r}^2
(\Delta) \rangle$, where $\delta \bm{r} (\Delta) \equiv \bm{r} (\Delta) - \bm{r}
(0)$, and the time-averaged mean square displacement (TMSD) $ \overline{\delta
  \bm{r}^2} (\Delta, t) $ which is defined below. Here, $\Delta$ is the lag
time, and $t$ is the total measurement time. The ensemble average $\left\langle
\dots \right\rangle$ is taken over both the thermal history and random
environment (realizations of random fractals and the quenched disorder of
traps). Using a method presented in \cite{meroz10, zumofen83, *blumen84}, we
have
\begin{align}
  \left\langle \delta \bm{r}^2 (\Delta) \right\rangle
  &\approx
  \int_{0}^{\infty} \left\langle \delta \bm{\tilde{r}}^2 (n) \right\rangle
  f_{\mu, \alpha} (n, \Delta) dn \notag\\[0.1cm]
  \label{e.eamsd}
  &\approx
  \int_{0}^{\infty} (x b_{\Delta})^{\beta}
  g_{\mu, \alpha} (x) dx
  \sim \Delta^{\mu \beta / \alpha},
\end{align}
where we used Eqs.~(\ref{e.eamsd.dtrw}), (\ref{e.rwsd.x_t}), (\ref{e.rwsd.x})
and (\ref{e.f-g}). Thus, EMSD shows anomalous subdiffusion [See
Fig.~\ref{f.eamsd}(a)]. Only when $\alpha=1$ (or, equivalently, $d_f = 2$), does
the scaling exponent of the subdiffusion coincide with that of the CTRW
\cite{meroz10}.

The TMSD, $\overline{\delta \bm{r}^2} (\Delta, t)$, is defined by
\cite{golding06, akimoto11, he08}
\begin{equation}
  \label{e.tamsd.1}
  \overline{\delta \bm{r}^2} (\Delta, t)
  =
  \frac {1}{t - \Delta} \int_{0}^{t-\Delta} dt'
  \left| \bm{r}(t'+\Delta) - \bm{r}(t')\right|^{2},
\end{equation}
This TMSD is often used in single-particle tracking experiments because it is
difficult in general to obtain many trajectories.
%
We rewrite the TMSD as \cite{miyaguchi11c, miyaguchi13}
\begin{equation}
  \label{e.tamsd.2}
  \overline{\delta \bm{r}^2} (\Delta, t)
  \approx \frac {1}{t}\sum_{k=1}^{N_t} H_k (\Delta),
\end{equation}
with
\begin{equation}
  H_k (\Delta) \equiv |\delta \bm{r}_k|^2 \Delta + 2 \sum_{l=1}^{k-1} (\delta \bm{r}_k
  \cdot \delta \bm{r}_l) \theta (\Delta - (T_k - T_l)),  
\end{equation}
where $\delta \bm{r}_k \in \mathbb{R}^d$ is the displacement at time $T_k$, and
$\theta(t)$ is defined by $\theta(t) = t$ for $t\geq 0$, otherwise $\theta (t) =
0$. These equations can be derived by expressing $\bm{r}(t')$ as $\bm{r}(t') =
\sum_{k=1}^{\infty} \delta \bm{r}_k I(T_k < t')$, where $I(\dots)$ is the
indicator function, i.e., $I (\dots) = 1$ if the inside of the bracket is
satisfied, while $I (\dots) = 0$ otherwise.  Then, expressing the integrand in
Eq.~(\ref{e.tamsd.1}), $\left| \bm{r}(t'+\Delta) - \bm{r}(t')\right|^{2}$, with
$\delta \bm{r}_k$ and the indicator function, we obtain Eq.~(\ref{e.tamsd.2}).

From Eq.~(\ref{e.tamsd.2}), we have
\begin{equation}
  \label{e.tamsd.3}
  \overline{\delta \bm{r}^2} (\Delta, t)
  \approx \frac {N_t}{t} \frac {1}{N_t}\sum_{k=1}^{N_t} H_k (\Delta)
  \to
  \frac {N_t}{t} h(\Delta)
\end{equation}
for large $t$. Here, we assume that the law of large numbers is satisfied for
the summation of the random variables $H_k (\Delta)$. This assumption can be
proved for hypercubic lattices \cite{miyaguchi11c} and confined systems
\cite{miyaguchi13}, whereas a general proof seems difficult, because the
correlation between displacements $\delta \bm{r}_k$ should be taken into
account. Nevertheless, this assumption is reasonable, because it is essentially
the ergodic hypothesis for the DTRW \cite{miyaguchi13}, and it is well accepted
fact that DTRWs on fractals are ergodic \cite{szymanski09}.  The important point
is that the statistical properties of the TMSD are completely determined by
$N_t$, and we have already derived the PDF of $N_t$ in Eqs.~(\ref{e.mod_ml}) and
(\ref{e.f-g}).


\begin{figure}[t]
  \centerline{\includegraphics[width=8.6cm]{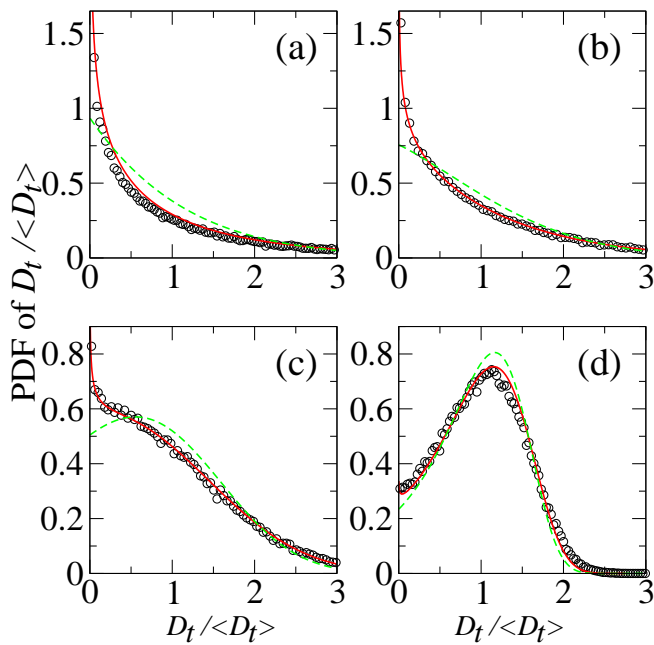}}
  \caption{\label{f.dist}(Color online) Rescaled PDF of the diffusion constant
    calculated simply by $D_t/\left\langle D_t \right\rangle = \overline{\delta
      \bm{r}^2} (\Delta, t) / \langle \overline{\delta \bm{r}^2} (\Delta, t)
    \rangle$ [Eq.~(\ref{e.tamsd.f})] for $\Delta = 10^5$ and $t= 2 \times
    10^{9}$. The parameter $\mu$ is set as (a) $\mu = 0.2$, (b) $\mu = 0.4$, (c)
    $\mu = 0.6$, and (d) $\mu = 0.8$. The solid curves are the theoretical
    result [Eq.~(\ref{e.mod_ml})], and the dashed curves are the MLD.  }
\end{figure}


The ensemble average of Eq.~(\ref{e.tamsd.3})  gives
\begin{equation}
  \label{e.ea-tamsd.1}
  \left\langle \overline{\delta \bm{r}^2} (\Delta, t) \right\rangle
  \sim
  t^{\frac {\mu}{\alpha} - 1} h(\Delta),
\end{equation}
where we used $\left\langle N_t \right\rangle \sim b_t \sim t^{\mu/\alpha}$.  On the other
hand, the ensemble average of Eq.~(\ref{e.tamsd.1}) gives
\begin{align}
  \left\langle \overline{\delta \bm{r}^2} (\Delta, t) \right\rangle
  &\sim
  \frac {1}{t} 
  \int_{0}^t dt'
  \left\langle \left| \bm{r}(t'+\Delta) - \bm{r}(t')\right|^{2} \right\rangle
  \notag \\[0.0cm]
  &\sim
  \frac {\Delta^{1 + \mu\beta/\alpha}}{t} 
  \int_{0}^{t/\Delta} dt' q(t')
  \label{e.ea-tamsd.2}
\end{align}
Here, we assumed a scaling hypothesis for the integrand $\langle\left|
\bm{r}(\Delta (t'+1)) - \bm{r}(\Delta t')\right|^{2} \rangle \sim
\Delta^{\gamma} q(t')$ with a constant $\gamma$ and an unknown function
$q(t')$. By setting $t' = 0$, we found $\gamma = \mu \beta /\alpha$ from
Eq.~(\ref{e.eamsd}). Note that on fractal lattices, $\langle\left| \bm{r}(\Delta
(t'+1)) - \bm{r}(\Delta t')\right|^{2} \rangle \neq \langle\bm{r}^2(\Delta
(t'+1))\rangle - \langle\bm{r}^2(\Delta t')\rangle$ in general, due to
correlations between successive jump directions.

Comparing Eqs.~(\ref{e.ea-tamsd.1}) and (\ref{e.ea-tamsd.2}), we have
$\int_{0}^{t/\Delta} dt' q(t') \sim \left(t/\Delta\right)^{\mu/\alpha}$, and
the ensemble-averaged TMSD (ETMSD) is given by
\begin{equation}
  \label{e.ea-tamsd.f}
  \left\langle \overline{\delta \bm{r}^2} (\Delta, t) \right\rangle
  \sim
  \frac {\Delta^{1 + (\beta-1)\mu/\alpha}}{t^{1 - \mu/\alpha}}.
\end{equation}
Thus, the ETMSD shows subdiffusion [$\Delta^{1 + (\beta-1)\mu/\alpha}$] as well
as aging ($1/t^{1 - \mu/\alpha}$). See also Figs.~\ref{f.eamsd}(b) and (c). Note
that the above formula is equivalent to that for CTRWs \cite{meroz10} if $\alpha
= 1$ (or equivalently, $d_f = 2$). By contrast, if $\alpha < 1$ (or
equivalently, $d_f < 2$), the above equation is not equivalent to that of CTRWs.


Finally, we derive the PDF of the generalized diffusion coefficient of the TMSD.
From Eqs.~(\ref{e.ea-tamsd.1}) and (\ref{e.ea-tamsd.f}), we obtain $h(\Delta) =
\Delta^{1 + (\beta-1)\mu/\alpha}$ and thus Eq.~(\ref{e.tamsd.3}) is rewritten as
\begin{equation}
  \label{e.tamsd.f}
  \overline{\delta \bm{r}^2} (\Delta, t)
  \sim
  \frac {N_t}t \Delta^{1 + (\beta-1)\mu/\alpha}.
\end{equation}
It follows that the generalized diffusion coefficient $D_t$ is given by
$D_t \sim N_t/t$
and therefore $D_t/\left\langle D_t \right\rangle$ follows the same PDF as
$N_t/\left\langle N_t \right\rangle$ (see Fig.~\ref{f.dist}). If the system is
ergodic, this PDF converges to a delta function, that is $D_t \to \left\langle
D_t \right\rangle$ as $t \to \infty$. However, this is not the case in the
present model; the PDF converges to the modified MLD $g_{\mu, \alpha} (x)$
[Eq.~(\ref{e.mod_ml})], and thus the ergodicity breaks down weakly with
everlasting randomness of time-averaged quantities \cite{he08, meroz10,
  miyaguchi11b, aaronson97, miyaguchi11c, miyaguchi13, rebenshtok07, *jeon13,
  *froemberg13, *froemberg13b, *cherstvy13, akimoto13, *akimoto14, Note1}.
%


In summary, the QTM on fractal lattices was investigated and anomalous
subdiffusion was found for both EMSD and TMSD.  It is also shown that this
system shows weak ergodicity breaking, and the diffusion constant of the TMSD
becomes a random variable following the modified MLD. This modified MLD has a
divergent peak at the origin, which means that there are trajectories with small
diffusivity much more frequently in the QTM than in the CTRW.

We also show that if the spectral dimension $d_s$ of the fractal lattice
satisfies $d_s < 2$, the QTM cannot be reduced to the CTRW; in other words, the
CTRW is physically irrelevant as a model of a random walk on random potential
energy landscapes and we have to use the QTM instead of the CTRW (though, if
$d_s$ is close to $2$, the CTRW is a good approximation of the QTM).  Only if
$d_s = 2$ the QTM is asymptotically equivalent to the CTRW.
Finally, it is worth mentioning that, even though we focused on the TMSD as a
time-averaged observable, the weak ergodicity breaking and the modified MLD must
appear for a wide class of observables \cite{miyaguchi11c, miyaguchi13}.




\end {document}